\documentclass{article}

\usepackage{microtype}
\usepackage{graphicx}
\usepackage{subfigure}
\usepackage{booktabs} % for professional tables

\usepackage{hyperref}

\usepackage[preprint]{icml2020}

\usepackage{amsmath}
\usepackage{amsthm}
\usepackage{amssymb}
\usepackage{mathtools}
\usepackage{vwcol}
\usepackage{multirow}
\usepackage{dcolumn}
\usepackage{url}
\usepackage{commath}
\usepackage{bm}
\usepackage{verbatim}

\newcommand{\bpi}{\bar{\pi}}
\newcommand{\ba}{\bar{a}}
\newcommand{\br}{\bar{r}}
\newcommand{\bA}{\bar{\mathcal{A}}}

\icmltitlerunning{Scalable Multi-Agent Inverse Reinforcement Learning via Actor-Attention-Critic}

\begin{document}

\twocolumn[
\icmltitle{Scalable Multi-Agent Inverse Reinforcement Learning via Actor-Attention-Critic}

% It is OKAY to include author information, even for blind
% submissions: the style file will automatically remove it for you
% unless you've provided the [accepted] option to the icml2020
% package.

% List of affiliations: The first argument should be a (short)
% identifier you will use later to specify author affiliations
% Academic affiliations should list Department, University, City, Region, Country
% Industry affiliations should list Company, City, Region, Country

% You can specify symbols, otherwise they are numbered in order.
% Ideally, you should not use this facility. Affiliations will be numbered
% in order of appearance and this is the preferred way.
% \icmlsetsymbol{equal}{*}

\begin{icmlauthorlist}
\icmlauthor{Wonseok Jeon}{Mila,McGill}
\icmlauthor{Paul Barde}{Mila,McGill}
\icmlauthor{Derek Nowrouzezahrai}{Mila,McGill}
\icmlauthor{Joelle Pineau}{Mila,McGill}
\end{icmlauthorlist}

\icmlaffiliation{Mila}{Mila, Quebec AI Institute}
\icmlaffiliation{McGill}{McGill University}

\icmlcorrespondingauthor{Wonseok Jeon}{jeonwons@mila.quebec}

% You may provide any keywords that you
% find helpful for describing your paper; these are used to populate
% the "keywords" metadata in the PDF but will not be shown in the document
\icmlkeywords{imitation learning, inverse reinforcement learning, multi-agent learning, scalability, sample efficiency}

\vskip 0.3in
]

% this must go after the closing bracket ] following \twocolumn[ ...

% This command actually creates the footnote in the first column
% listing the affiliations and the copyright notice.
% The command takes one argument, which is text to display at the start of the footnote.
% The \icmlEqualContribution command is standard text for equal contribution.
% Remove it (just {}) if you do not need this facility.

\printAffiliationsAndNotice{}  % leave blank if no need to mention equal contribution
%\printAffiliationsAndNotice{\icmlEqualContribution} % otherwise use the standard text.

%%%%%%%%%%%%%%%% NOW, DO NOT MODIFY ABSTRACT!!!!
\begin{abstract}
Multi-agent adversarial inverse reinforcement learning (MA-AIRL) is a recent approach that applies single-agent AIRL to multi-agent problems where we seek to recover both policies for our agents and reward functions that promote expert-like behavior. While MA-AIRL has promising results on cooperative and competitive tasks, it is sample-inefficient and has only been validated empirically for small numbers of agents -- its ability to scale to many agents remains an open question. We propose a multi-agent inverse RL algorithm that is more sample-efficient and scalable than previous works. Specifically, we employ multi-agent actor-attention-critic (MAAC) -- an off-policy multi-agent RL (MARL) method -- for the RL inner loop of the inverse RL procedure. In doing so, we are able to increase sample efficiency compared to state-of-the-art baselines, across both small- and large-scale tasks. Moreover, the RL agents trained on the rewards recovered by our method better match the experts than those trained on the rewards derived from the baselines. Finally, our method requires far fewer agent-environment interactions, particularly as the number of agents increases.
\end{abstract}

%%%%%%%%%%%%%%%%%%%%%%
\section{Introduction}

Inverse reinforcement learning (IRL)~\cite{ng2000algorithms} captures the problem of inferring a reward function that reflects the objective function of an expert, from limited observations of the expert's behavior.
Traditionally, IRL required planning algorithms as an inner step~\cite{ziebart2008maximum}, which makes IRL expensive in high-dimensional control tasks.
Later work alleviates this by using adversarial training objectives~\cite{ho2016generative,fu2018learning}, inspired by the generative adversarial network (GAN) method~\cite{goodfellow2014generative}.
Essentially, these methods iteratively train a discriminator to measure the difference between the agent's and the expert's behaviors, and optimize a policy to reduce such difference via reinforcement learning.
Combined with modern deep RL algorithms~\cite{schulman2015trust}, adversarial imitation and IRL show improved scalability to high-dimensional tasks.

Recently, adversarial imitation and IRL have been extended to multi-agent imitation~\cite{song2018multi} and multi-agent IRL~\cite{yu2019multi}, respectively, where the agents in the same environment aim to learn rewards or policies from multiple experts' demonstrations.
Both these previous works have shown strong theoretical relationship between single-agent and multi-agent learning methods, and demonstrated that the proposed methods outperform baselines.
However, there remains some questions on their empirical performances.
First of all, both~\citet{song2018multi} and \citet{yu2019multi} have focused on the  performance \textit{after} convergence, but the sample efficiency of the proposed methods in terms of agent-environment interactions has not been considered rigorously. 
Also, both used the multi-agent extension of ACKTR, MACK~\cite{wu2017scalable}, which is built upon the \emph{centralized-training decentralized-execution}  framework~\cite{lowe2017multi,foerster2018counterfactual} and uses centralized critics to stabilize training.
If such centralized critics are not carefully designed, the resultant MARL algorithm may scale poorly with the number of agents due to the curse of dimensionality, i.e., the joint observation-action space grows exponentially with the number of agents.

In this work, we propose multi-agent discriminator-actor-attention-critic (MA-DAAC), a multi-agent algorithm capable of sample-efficient imitation and inverse reward learning, regardless of the number of agents.
MA-DAAC uses multi-agent actor-attention-critic~\cite{iqbal2019actor} that scales to large numbers of agents thanks to a shared attention-critic network~\cite{vaswani2017attention}.
We verify that MA-DAAC is more sample-efficient than the multi-agent imitation and IRL baselines, and demonstrate that the reward functions learned by MA-DAAC lead to improved empirical performances.
Finally, MA-DAAC is shown to be more robust to smaller number of experts' demonstration.

%%%%%%%%%%%%%%%%%%%%%%%%%%%%%%%%%%%%%%%%%%%%%%
\section{Preliminaries\vspace{-0.3in}}~\label{sec:background}

%%%%%%%%%%%%%%%%%%%%%%%%%%%%%%%%%%%%%%%
\subsection{Markov Games and Notations}
In a Markov Game~\cite{littman1994markov}, multiple agents observe a shared environment state and take individual actions based on their observations. 
Then, each agent gets a reward and the environment transitions to a new state.
Mathematically, a Markov Game is defined by 
$\langle\mathcal{S}, \{\mathcal{A}_i\}_{i=1}^N, \{r_i\}_{i=1}^N, P_T, \nu, \gamma\rangle$, where
$N$ is the number of agents,
$\mathcal{S}$ is the state space, 
$\mathcal{A}_i$ is the action space for agent $i$, 
$r_i(s, a_1, ..., a_N)$ is a reward function for agent $i$, 
$P_T(s'|s, a_1, ..., a_N)$ is a state transition distribution, 
$\nu(s)$ is an initial state distribution,
$\gamma$ is a discount factor. 
Also, the policy $\pi_i(a_i|s)$ is the probability of the $i$-th agent choosing an action $a_i$ at the state $s$.
For succinct notation, we use bars to indicate joint quantities over the agents, e.g.,
$\bA=\mathcal{A}_1\times\cdots\mathcal{A}_N$ is a joint action space,
$\ba=(a_1, ..., a_N)$ is a joint action,
$\bpi=(\pi_1, ..., \pi_N)$ is a joint policy,
$\br(s, \ba)=(r_1(s, \ba), r_2(s, \ba), ..., r_N(s, \ba))$ is a joint reward.
The value function of the $i$-th agent with respect to $\bpi$ is defined by
$
Q_i^{\bpi}(s, \ba)=
\mathbb{E}^{\bpi}
[
{\textstyle \sum_{t=0}^\infty}
\gamma^t
r_i(s_t, \ba_t)
|
s_0=s,
\ba_0=\ba
]
$,
where the superscript $\bpi$ on the expectation implies that states and joint actions are sampled from $\nu, P_T$ and $\bpi$.
Additionally, the $\gamma$-discounted state occupancy measure $\rho^{\bpi}$ of the joint policy $\bpi$ is defined by
$
\rho^{\bpi}(s, \ba)
=
\mathbb{E}^{\bpi}
[
\sum_{t=0}^\infty
\gamma^t
\mathbb{I}\{s_t=s, \ba_t=\ba\}
]
$
,
where $\mathbb{I}\{\cdot\}$ is an indicator function.

In this work, we consider a partially observable Markov Game, 
where each agent can only use its own observation $o_i\in\mathcal{O}_i$ from the environment's state $s\in\mathcal{S}$.
Due to the partial observability, we consider the policy $\pi_i(a_i|o_i)$ which is the probability that the $i$-th agent chooses an action $a_i$ after observing $o_i$.
Also, we consider value functions of the form $Q_i^{\bpi}(\bar{o}, \ba)$ that use the joint observation $\bar{o}:=(o_1, ..., o_N)$ instead of the state $s$, which is commonly done in previous works~\cite{lowe2017multi,iqbal2019actor}.

%%%%%%%%%%%%%%%%%%%%%%%%%%%%%%%%%%%%%%%%%%%%%%%%%%%%%%%
\subsection{Multi-Agent Adversarial Imitation and IRL}

In the multi-agent IRL problem, $N$ agents respectively try to mimic $N$ experts' policies  $\bpi^E=(\pi_{1}^E, ..., \pi_{N}^E)$.
Here, each agent is not allowed to access to its own target expert's policy directly and should rely on a limited amount of experts' demonstration.
There are two possible objectives in this problem; (1) policy imitation -- learning policies close to those of the experts -- and (2) reward learning -- recovering reward functions that lead to expert-like behavior.

Multi-agent generative adversarial imitation learning (MA-GAIL) enables agents to learn experts' policies by optimizing the following mini-max objective~\cite{song2018multi}:
\begin{align*}
\min_{\bpi}
\max_{D_1,\ldots,D_N}
&\mathbb{E}_{s, \ba\sim\rho^{\bpi}}
\left[
{\textstyle\sum_{i=1}^N}
\log D_i(s, a_i)
\right]\\
&+
\mathbb{E}_{s, \ba\sim\rho^{\bpi^E}}
\left[
{\textstyle\sum_{i=1}^N}
\log
(1-D_i(s, a_i))
\right].
\end{align*}
In practice, MA-GAIL iteratively optimizes discriminators $D_1, \ldots , D_N$ and policies $\bpi$, where the discriminators are trained to classify whether state-action pairs come from agents or experts and the policies are optimized with MARL methods and reward functions $\log D_i(s, a_i)$ defined by the discriminators. Although MA-GAIL successfully imitates experts, learned rewards $\log D_i(s, a_i)$ of MA-GAIL cannot be used as reward functions due to discriminators converging to $1/2$~\cite{goodfellow2014generative, fu2018learning}.

MA-AIRL addressed this issue by modifying the following parts in MA-GAIL.
First, structured form of discriminators motivated by logistic stochastic best response equilibrium (LSBRE) was used~\cite{yu2019multi}:
\begin{align*}
    D_i(s, a_i, s')&=\frac{\exp(f_i(s, a_i, s'))}{\exp(f_i(s, a_i, s'))+\pi_i(a_i|s)},\\
    f_i(s, a_i, s')&=g(s, a_i)+\gamma h(s') - h(s).
\end{align*}
In addition, MA-AIRL used $\log D_i(s, a_i, s') - \log (1 - D_i(s, a_i, s'))$ as reward functions instead of the functions $\log D_i(s, a_i)$ of MA-GAIL.
It turns out that either $f_i$ or $g$ can recover the reward functions that lead to the experts' behavior.

%%%%%%%%%%%%%%%%%%%%%%%
\section{Related Works}

%%%%%%%%%%%%%%%%%%%%%%%%%%%%%%%%%%%%%%%%%%%%%%%%%%%%%%%
\subsection{Sample-Efficient Adversarial Imitation Learning}

In~\citet{kostrikov2018discriminator}, TD3~\cite{fujimoto2018addressing} was used with a discriminator.
To stabilize their algorithm, they proposed to learn the terminal-state values using a discriminator and use them in the RL inner loop of imitation learning, whereas conventional off-policy reinforcement learning algorithms implicitly consider zero terminal-state values and do not account for them. 
In~\citet{sasaki2018sample}, another sample-efficient imitation learning algorithm was proposed. 
In contrast with prior works, their method did not use discriminators by considering the Bellman consistency of the imitation learning reward signal. 
Then, by using off-policy actor-critic~\cite{degris2012off}, it was shown that the proposed method is much more sample-efficient than GAIL.

%%%%%%%%%%%%%%%%%%%%%%%%%%%%%%%%%%%%%%%%%%
\subsection{Scalable Multi-Agent Learning}

For large number of agents,
it has been regarded as a challenging problem for MARL to achieve coordinated multi-agent behavior.
Although existing works using centralized critics such as MADDPG~\cite{lowe2017multi} and COMA~\cite{foerster2018counterfactual} make a handful of agents coordinated, they struggle when the number of agents to manage increases.
This is mainly due to the exponential growth of the critic inputs with the increasing number of agents, which possibly increases the input variance of training as well.
MAAC~\cite{iqbal2019actor} addressed such an issue by using the attention mechanism~\cite{vaswani2017attention} and a shared critic network, and it outperformed existing algorithms for large number of agents.
Thanks to the attention mechanism, MAAC is trained to focus only on part of joint observations and actions, which leads to rapid and efficient training.
Mean-field MARL~\cite{yang2018mean} is another approach to address the scalability issue of MARL, which is based on mean-field approximation for the centralized critics.
However, its application is restricted to the situation where all agents are homogeneous, whereas MAAC can be applied to much general scenarios in which non-homogeneous agents exist.

Meanwhile, there were some multi-agent imitation learning algorithms applied to large-scale environments.
\citet{le2017coordinated} proposed multi-agent imitation learning in the index-free control setting, where agents are not allowed to know the indices of their own expert. 
The proposed method trains a model that infers and assigns the role of each agent with rollout trajectories and given experts' demonstration and exploits that model to make highly coordinated behavior.
\citet{sanghvi2019modeling} uses multi-agent imitation learning to learn social group communication among agents with a single shared policy network among agents.
However, they used multi-agent behavioral cloning and focused on the environments with homogeneous agents.
In contrast with those works, \textit{our algorithm can deal with non-homogeneous agents as well}.
In addition, it has been reported in existing literature~\cite{kostrikov2018discriminator,sasaki2018sample,song2018multi} that behavioral cloning performs poorly when there are only small number of experts' demonstration due to the co-variate shift problem~\cite{ross2011reduction}.
For these reasons, we narrow down our scope to MA-GAIL and MA-AIRL in this work.

\begin{algorithm}[t]
    \caption{Multi-Agent Discriminator-Actor-Attention-Critic (MA-DAAC)}
\label{algorithm:MAA2CIL}
\begin{algorithmic}[1]

    \STATE {\bfseries Input:}
    a buffer $\mathcal{B}_A$ for agents' rollout trajectories,
    experts' demonstration $\mathcal{B}_E$,
    policy networks $\bar{\pi}_{\bar{\theta}}=\{\pi_{\theta_i}\}_{i=1}^N$,
    a shared attention critic $\bar{Q}_{\bar{\phi}}=\{Q_{\phi_i}\}_{i=1}^N$, and
    discriminators $\bar{D}_{\bar{\omega},\bar{\psi}}=\{D_{\omega_i,\psi_i}\}_{i=1}^N$
    
    \FOR{each iteration}
    
        \STATE Sample rollout trajectories:
                $
                (\bar{o}, \bar{a}, \bar{o}')\sim \bar{\pi}_{\bar{\theta}}.
                $

        \STATE Add sampled trajectories to $\mathcal{B}_A$.
    
        \FOR{each training iteration}
        
            \STATE Sample $(\bar{o}^A, \bar{a}^A, \bar{o}'{}^A)$ from $\mathcal{B}_A$.
            
            \STATE Sample $(\bar{o}^E, \bar{a}^E, \bar{o}'{}^E)$ from $\mathcal{B}_E$.
                
            \STATE Update rewards by using discriminators.
                \begin{align*}
                    r_i^A(\bar{o}^A, \bar{a}^A, \bar{o}'{}^A)&=
                    \log D_{\omega_i,\psi_i}(\bar{o}^A, \bar{a}^A, \bar{o}'{}^A)\\
                    &-
                    \log (1 - D_{\omega_i,\psi_i}(\bar{o}^A, \bar{a}^A, \bar{o}'{}^A))
                \end{align*}
            \STATE 
            // Policy learning via MAAC
            
            Update $\bar{\theta}, \bar{\phi}$ with \eqref{eq:critic} and \eqref{eq:policy}.
            
            \STATE 
            // Reward learning
            
            Update $\bar{\omega},\bar{\psi}$ with \eqref{eq:disc_op}.
            
        \ENDFOR
    \ENDFOR
    
    \STATE {\bfseries Output:} $\bar{\theta}, \bar{\omega}$
\end{algorithmic}
\end{algorithm}

%%%%%%%%%%%%%%%%%%%%
\section{Our Method\vspace{-0.2in}}~\label{sec:ourapproach}

We consider multi-agent IRL problems where agents desire to learn experts' behavior as well as reward functions that lead to such behavior.
Note that agents cannot access to experts directly, but learning agents are allowed to used the limited amount of experts' demonstration.
In such setting, we introduce MA-DAAC, our multi-agent IRL method as outlined in \textbf{Algorithm~\ref{algorithm:MAA2CIL}}.
Our method iteratively trains discriminators and policies using experts' demonstration and agents' rollout trajectories -- using MARL and a reward signal modeled by the discriminators -- similar to MA-GAIL and MA-AIRL.

%%%%%%%%%%%%%%%%%%%%%%%%
\textbf{MARL Algorithm.} 
In our method, we use MAAC, which are off-policy MARL algorithms and shown to be sample-efficient and scalable to large number of agents~\cite{iqbal2019actor}.
We summarize it as follows.
Assuming discrete action spaces, let  
\begin{align*}
    \bar{\vec{Q}}_{\bar{\phi}}(\bar{o}, \bar{a})
    =
    (\vec{Q}_{\phi_1}(\bar{o}, \bar{a}), ..., \vec{Q}_{\phi_N}(\bar{o}, \bar{a})),
\end{align*}
denote action values, where each element of $\bar{\vec{Q}}_{\bar{\phi}}(\bar{o}, \bar{a})$ is a vector-valued action value of the corresponding agents and $\bar{\phi}=(\phi_1, ..., \phi_N)$ is a set of neural network parameters for the critic network.
In MAAC, the $i$-th agent's action value was modeled as a neural network
\begin{align*}
    \vec{Q}_{\phi_i}(\bar{o}, \bar{a})
    =
    \vec{f}_i(
    g_i^{\mbox{\scriptsize{local}}}(o_i), 
    g_i^{\mbox{\scriptsize{global}}}(\bar{o}, \bar{a})
    )
    ,
\end{align*}
where $g_i^{\mbox{\scriptsize{local}}}$ is a network looking at the $i$-th agent's local observation and action, $g_i^{\mbox{\scriptsize{global}}}$ is another network that considers \emph{other} agents' observations and actions. Finally, $\vec{f}_i$ is a network that takes into account the extracted features from both of the previous neural networks.
The main idea of MAAC is to model $g_i^{\mbox{\scriptsize{global}}}$ with an attention mechanism~\cite{vaswani2017attention} and to share this network among agents, i.e., for the $i$-th agent's embedding $e_i$, 
\begin{align*}
    g_i^{\mbox{\scriptsize{global}}}(\bar{o}, \bar{a})
    =
    \sum_{j\ne i}
    P_{i\rightarrow j}(o_i, a_i, o_j, a_j)
    v_j(o_j, a_j),
\end{align*}
where
\begin{align*}
    P_{i\rightarrow j}(o_i, a_i, o_j, a_j)
    &\propto 
    \exp(
    (W_Ke_j(o_j, a_j))^T
    W_Qe_i(o_i, a_i)
    ),\\
    v_j(o_j, a_j)
    &=
    \sigma(W_Ve_j(o_j, a_j))
\end{align*}
for element-wise non-linear activation $\sigma$ and shared neural network parameters $W_K$, $W_Q$ and $W_V$ among agents. The objective of the critic training is to minimize the sum of temporal difference (TD) errors:
\begin{align}
    \underset{\bar{\phi}}{\mathrm{argmin}}~
    \mathbb{E}_{\bar{o}, \bar{a}\sim\rho_{\mathcal{B}}}
    \left[
    {\textstyle \sum_{i=1}^N }
    (
        y_i(\bar{o}, \bar{a}, \bar{o}')
        -
        Q_{\phi_i}(\bar{o}, \bar{a})
    )^2
    \right],\label{eq:critic}
\end{align} 
where $
    y_i(\bar{o}, \bar{a}, \bar{o}')
    =
    r_i(\bar{o}, \bar{a}) 
    +
    \gamma 
    \mathbb{E}_{\bar{a}'\sim\bar{\pi}_{\bar{\theta}'}(\cdot|\bar{o}')}
    Q_{\phi_i'}(\bar{o}', \bar{a}')
$, $\bar{o}, \bar{a}\sim\rho_{\mathcal{B}}$ implies $\bar{o}, \bar{a}$ are sampled from an experience replay buffer $\mathcal{B}$, $Q_{\phi_i}$ is the value for $a_i$ in $\vec{Q}_{\phi_i}$, $\bar{\theta}'$ is a target policy parameter, $\phi_i'$ is a target critic parameter, and $r_i$ is the $i$-th agent's reward function.
For policy updates, the policy gradient
\begin{align}
    &\mathbb{E}_{
        \bar{o}\sim\rho_{\mathcal{B}},
        \bar{a}\sim\bar{\pi}_{\bar{\theta}}(\cdot|\bar{o})
    }
    \nabla_{\theta_i}
    \log\pi_{\theta_i}(a_i|o_i)
    A_i(\bar{o}, \bar{a}), \label{eq:policy}\\
    &A_i(\bar{o}, \bar{a})
    =
    Q_{\phi_i}(\bar{o}, \bar{a})
    -
    \sum_{a_i'\in\mathcal{A}_i}
    \pi_i(a_i'|o_i)
    Q_{\phi_i}(\bar{o}, \bar{a}_{-i}(a_i')) \nonumber
\end{align}
was used, where $\bar{a}_{-i}(a_i')$ is the change of the $i$-th action in $\bar{a}$ to $a_i'$.

%%%%%%%%%%%%%%%%%%%%%%%
\textbf{Discriminator.}
We consider two types of discriminator models that were considered in MA-GAIL~\cite{song2018multi}; a \emph{centralized discriminator} that takes all agents' observations and actions as its input and outputs multi-head classification for each agent; a \emph{decentralized discriminator} that takes each agent's local observations and actions and outputs a single-head classification result.
It should be noted that both sample efficiency and scalability for both discriminators have not been rigorously analyzed in MA-GAIL and MA-AIRL~\cite{song2018multi,yu2019multi}.

Let $
    \bar{D}(\bar{o}, \bar{a}, \bar{o}')=(D_1(\bar{o}, \bar{a}, \bar{o}'), ..., D_N(\bar{o}, \bar{a}, \bar{o}'))
$
denote the vector-valued discriminator output. 
This is a general expression for both types of discriminators since one can consider a shared feature among $D_1, ..., D_N$
for the centralized discriminator, while decentralized discriminators do not share feature but ignore other agents' observations and actions, i.e., $D_i(o_i, a_i, o_i')$.
For each training iteration, we train discriminator by maximizing
\begin{align}
    &\mathbb{E}_{\bar{o}^A, \bar{a}^A\sim\rho_{\bar{\pi}^A}}
    \left[
    {\textstyle \sum_{i=1}^N}
        \log (1 - D_i(\bar{o}^A, \bar{a}^A, \bar{o}'{}^A))
    \right]
    \nonumber
    \\
    &+
    \mathbb{E}_{\bar{o}^E, \bar{a}^E\sim\rho_{\bar{\pi}^E}}
    \left[
    {\textstyle \sum_{i=1}^N}
        \log D_i(\bar{o}^E, \bar{a}^E, \bar{o}'{}^E)
    \right].
    \label{eq:disc_op}
\end{align}
Intuitively, discriminators are trained in a way that expert-like behavior gets higher values, whereas non-expert-like behavior results in lower values.
Also, note that the objective $\eqref{eq:disc_op}$ means the use of rollout trajectories sampled from agents' policies $\bar{\pi}^A$ in the first expectation.
In practice, however, we use the samples from the replay buffer \emph{without} off-policy correction to enhance the sample-efficiency of discriminator training via sample reuse.
As shown in our experiments, such an abuse of samples does not harm the performance, which is similar to the results in~\citet{kostrikov2018discriminator}. 
Similar to MA-AIRL, we assume the discriminators are parameterized neural networks such that
\begin{align}
    D_{\omega_i, \psi_i}(\bar{o}, \bar{a}, \bar{o}')&=\frac{\exp(f_{\omega_i, \psi_i}(\bar{o}, \bar{a}, \bar{o}'))}{\exp(f_{\omega_i, \psi_i}(\bar{o}, \bar{a}, \bar{o}'))+\pi_i(a_i|o_i)},\nonumber\\
    f_{\omega_i, \psi_i}(\bar{o}, \bar{a}, \bar{o}')&=g_{\omega_i}(\bar{o}, \bar{a})+\gamma h_{\psi_i}(\bar{o}') - h_{\psi_i}(\bar{o}).\label{eq:reward}
\end{align}
During training, we use $\log D_i(\bar{o}, \bar{a}, \bar{o}') - \log (1 - D_i(\bar{o}, \bar{a}, \bar{o}'))$ for $i=1, ..., N$ as agents' reward functions. Especially for centralized discriminators, we use \emph{observation-only} rewards~\cite{fu2018learning} that ignore the action inputs of the discriminators and lead to slightly better performance than those using action inputs. In Section~\ref{sec:discussion}, we discuss about it in detail.

\begin{figure*}[t]
\centering
\includegraphics[width=\textwidth]{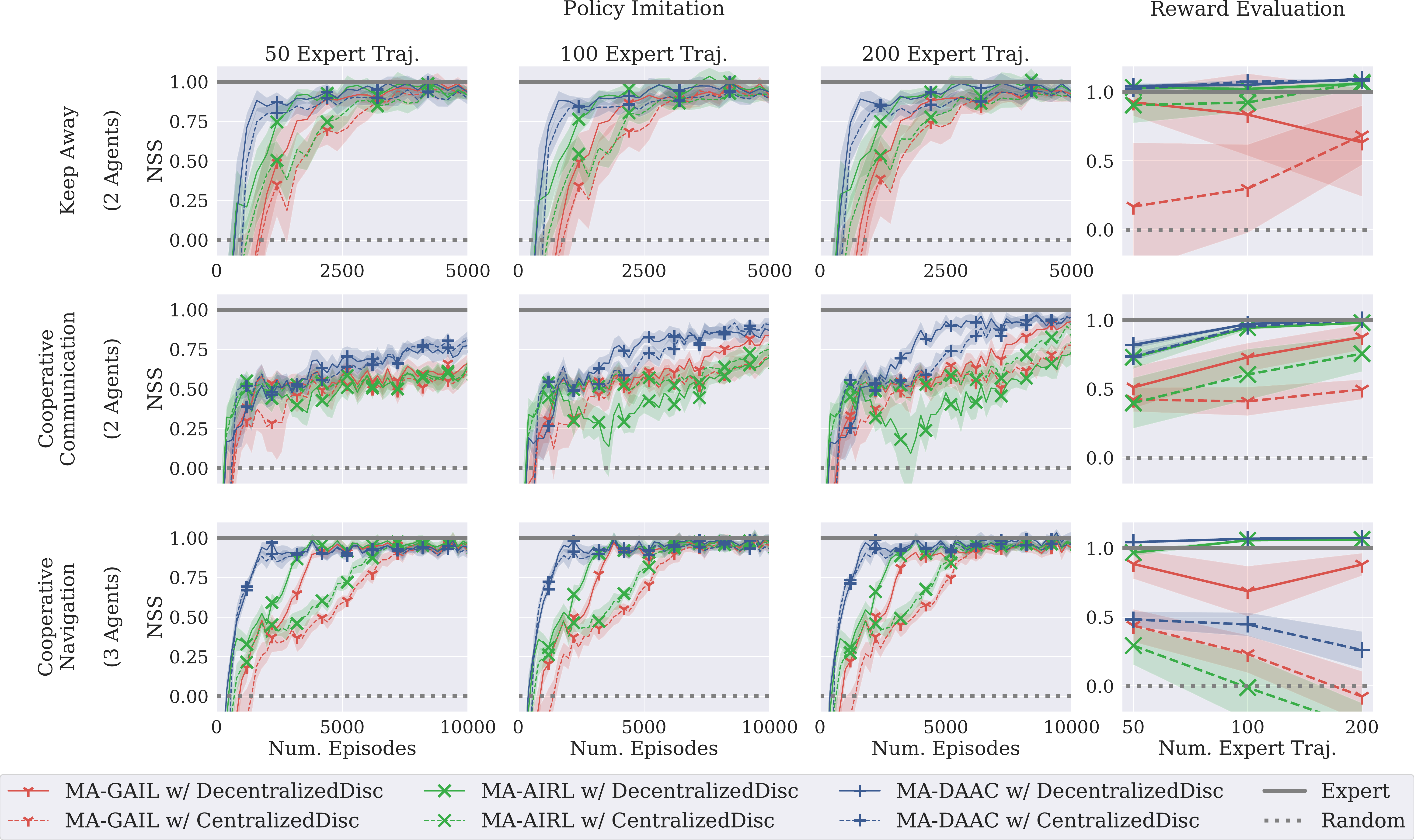}
\caption{
Imitation performance and reward learning performance in small-scale environments.
The results in the same row share their environments; \emph{Keey Away}, \emph{Cooperative Communication}, \emph{Cooperative Navigation} from top to bottom row.
For the first three columns, we report $\mathrm{NSS}$ of policies during training, where 50, 100 and 200 experts' demonstration were used from left to right column. Note that MA-DAAC converges to the best performance among all methods sample-efficiently.
For the last column, we report $\mathrm{NSS}$ of policies learned by reward functions from MA-DAAC. The result show that MA-DAAC always recover better reward functions than the baselines. Note that means and 95\% confidence intervals over 10 runs are considered.
}
\label{fig:smallscale}
\end{figure*}

\begin{figure*}[t]
\centering
\includegraphics[width=\textwidth]{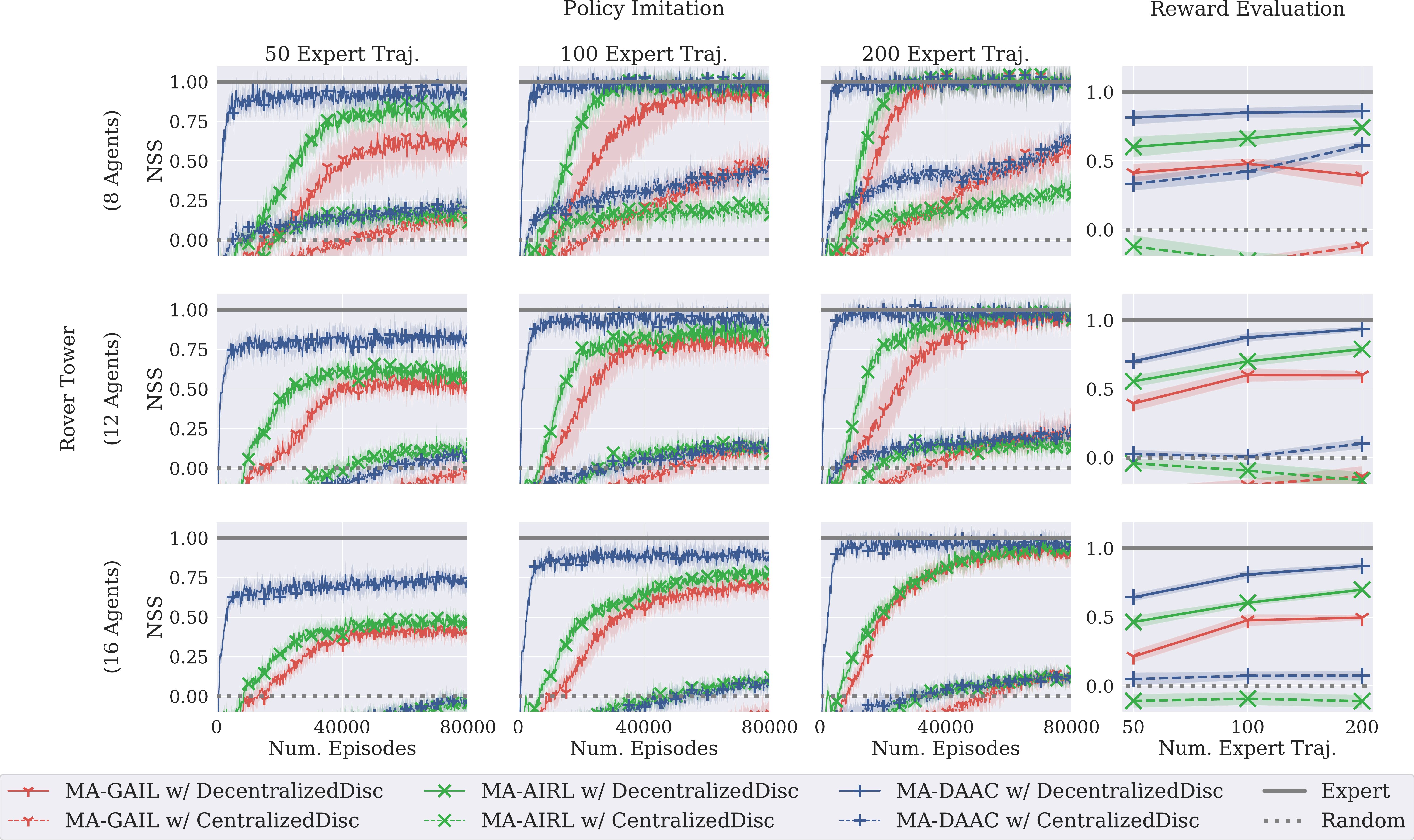}
\caption{
Imitation performance and reward learning performance in \emph{Rover Tower} tasks.
The results in the same row share the number of agents; 8, 12 and 16 from top to bottom row.
For the first three columns, we report $\mathrm{NSS}$ of policies during training, where 50, 100 and 200 experts' demonstration were used from left to right column.
Regardless of the number of agents, MA-DAAC converges much faster than the baselines. Also, its convergence is not affected much by the number of agents, whereas those of baselines becomes slower for the increasing number of agents. 
For the last column, we report $\mathrm{NSS}$ of policies learned by reward functions from MA-DAAC.
The policies learned by the rewards from MA-DAAC achieve higher $\mathrm{NSS}$. 
Note that means and 99\% confidence intervals over 10 runs are considered.
}
\label{fig:largescale}
\end{figure*}

%%%%%%%%%%%%%%%%%%%%%%%%%%%%%%%%%%%%%%%%%%%%%
\section{Experiments\vspace{-0.2in}}~\label{sec:experiments}

Our experiments are designed to answer the following questions:

\emph{1. Is MA-DAAC capable of recovering multi-agent reward functions effectively?}
\newline
\emph{2. Is MA-DAAC sample-efficient in terms of the number of agent-environment interactions and available expert demonstration?}
\newline
\emph{3. Is MA-DAAC scalable to the environments with many of agents?}

We evaluate our methods from two perspectives; \emph{policy imitation} and \emph{reward learning}. We briefly summarize our experiment setup in the following sections and include more detailed information in Appendix.

%%%%%%%%%%%%%%%%%%%%%%%%%%%%%
\subsection{Experiment Setup}

%%%%%%%%%%%%%%%
\textbf{Tasks.}
We consider two classes of environments, which respectively cover small-scale and large-scale environments. All of them run on OpenAI Multi-agent Particle Environment (MPE)~\cite{mordatch2017emergence}.
The small-scale environments include:

\emph{Keep Away $-$} There are 2 agents ``reacher" and ``pusher", where reacher tries to reach the goal and pusher tries to push it away from the goal.

\emph{Cooperative Communication $-$} There are 2 agents, ``speaker" and ``listener". One of three landmarks is randomly chosen as a target at each episode and its location can only be seen by speaker. Speaker cannot move, whereas listener can observe speaker's message and move toward the target landmark. 

\emph{Cooperative Navigation $-$} There are 3 agents and 3 landmarks, and the goal of agents is to cover as many landmarks as possible. 

We also consider a large-scale environments proposed in MAAC~\cite{iqbal2019actor} to measure the scalability of MA-DAAC and the existing methods:

\emph{Rover Tower~\cite{iqbal2019actor} $-$} There are even number of agents (8, 12, 16), where half of them are ``rovers" and the others are ``towers". For each episode, rovers and towers are \emph{randomly paired}, and each tower has its own goal destination. Similar to \textit{Cooperative Communication}, towers cannot move but can communicate with rovers so that rovers can move toward corresponding goals.

%%%%%%%%%%%%%%%%%
\textbf{Experts.}
For experts' policies, we trained policies by using MAAC over either 50,000 episodes (\emph{Keep Away}, \emph{Cooperative Communication}, \emph{Cooperative Navigation}) or 100,000 episodes (\emph{Rover Tower}). 
Then, we considered those trained policies as experts and generated trajectories from them, where the actions in those episodes are always taken with the largest probability.
Throughout our experiment, we vary the number of available demonstration from 50, 100 to 200.

%%%%%%%%%%%%%%%%%%%%%%%%%%%%%%%
\textbf{Performance Measure.}
In multi-agent IRL problems, we need to define a proper performance measure to see the gap between learned agents and experts during training. 
However, episodic-score-based measure widely used in single-agent learning~\cite{ho2016generative, kostrikov2018discriminator, sasaki2018sample} cannot be directly used in our problems due to the multiple reward functions for each agent and their \emph{unnormalized} scales.
Therefore, we define the \emph{normalized score similarity} ($\mathrm{NSS}$) as follows:
\begin{align*}
    \mathrm{NSS}
    =
    \frac{1}{N}\sum_{i=1}^N 
    \frac{{\mbox{score}}_{A_i} - {\mbox{score}}_{R_i}}{{\mbox{score}}_{E_i} - {\mbox{score}}_{R_i}}.
\end{align*}
Here, ${\mbox{score}}_{A_i}$ is the $i$-th agent's (episode) score during training, ${\mbox{score}}_{E_i}$ is the $i$-th expert's \emph{average} score for experts' demonstration, and ${\mbox{score}}_{R_i}$ is the \emph{average} score of the $i$-th agent when uniformly random actions were taken by all agents. 
Intuitively, $\mathrm{NSS}$ gets close to 1 if every agent shows expert-like behavior since such behavior will lead to the experts' score. 
One advantage of $\mathrm{NSS}$ is that we can evaluate multi-agent imitation performance for both competitive and cooperative tasks.
In our experiments, we show that it is an effective measure.

%%%%%%%%%%%%%%%%%%%%%%%%%%%%%%%%%%%%%
\subsection{Small-Scale Environments}

%%%%%%%%%%%%%%%%%%%%%%%%%%
\textbf{Policy Imitation.}
The results in the small-scale environments are summarized in \emph{Figure~\ref{fig:smallscale}} (column 1-3). 
For all small-scale environments, we demonstrate MA-DAAC converges faster than the baselines. 
This is due to the use of MAAC -- an off-policy MARL methods -- rather than using MACK -- an on-policy MARL methods -- proposed by~\citet{song2018multi}.
Also, there is only a negligible gap between the performances of using centralized discriminators and using decentralized discriminators.
It should be noted that in \citet{song2018multi}, imitation learning with centralized discriminators leads to slightly better performance compared to its decentralized counterparts, whereas we get comparable results for both types of discriminators. We believe this small difference comes from using different MACK implementations and experts' demonstration.

%%%%%%%%%%%%%%%%%%%%%%%%%
\textbf{Reward Learning.}
For all imitation and IRL methods, we first train both policies and rewards over 50,000 episodes and \emph{re-train} policies with MACK and learned rewards over 50,000 episodes. 
For rewards from MA-DAAC and MA-AIRL, we use $g_{\omega_i}$ in~\eqref{eq:reward}, a learned reward without potential functions $h_{\psi_i}$~\cite{fu2018learning,yu2019multi}.
For MA-GAIL, we use $\log D_i(\bar{o}, \bar{a})$ as compared by~\citet{yu2019multi}.
The results of reward learning are described in \emph{Figure~\ref{fig:smallscale}} (column 4).
Note that the policies trained by either imitation or IRL methods have comparable mean $\mathrm{NSS}$ for given the number of expert demonstration and environment. 
Nevertheless, the rewards learned by either MA-DAAC or MA-AIRL with decentralized discriminators achieve the best retraining performance. 
One interesting observation is that rewards from MA-DAAC with centralized discriminators lead to better performance compared to those from MA-AIRL with centralized discriminators. 
We believe using the off-policy samples in MA-DAAC makes the reward training more robust since MA-AIRL is trained by on-policy samples and can easily overfit to the latest rollouts. Additional results in small-scale environments are given in Appendix.

\begin{figure}[t]
\centering
\includegraphics[width=0.8\linewidth]{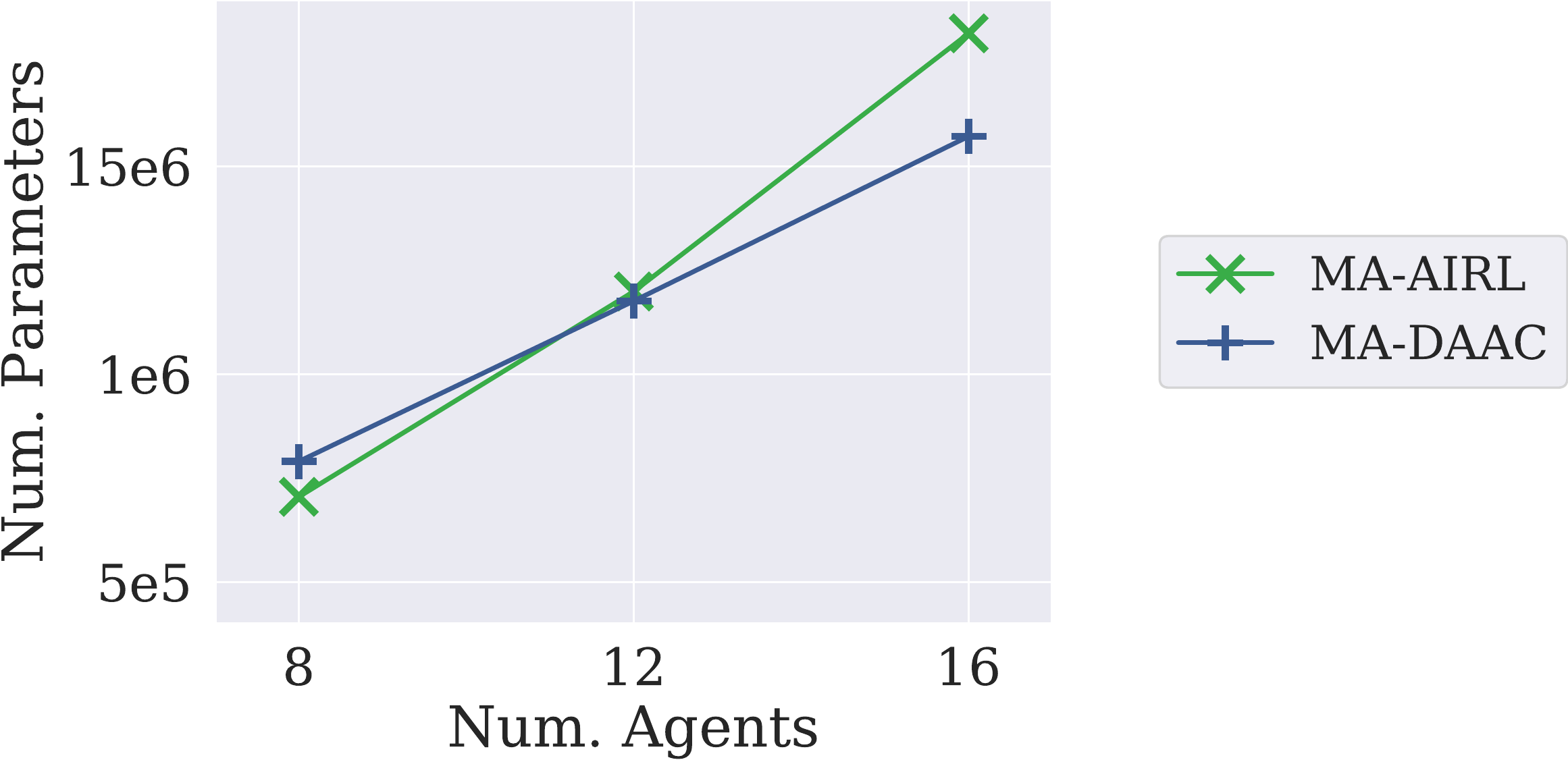}
\caption{The number of trainable parameters for each method with decentralized discriminator in \emph{Rover Tower} tasks. The number of parameters linearly increases for MA-DAAC, whereas it increases for MA-AIRL. Note that MA-DAAC performs better than MA-AIRL for both policy imitation and reward learning with fewer number of parameters in the environments with either 12 or 16 agents.}
\label{fig:largescale2}
\end{figure}

%%%%%%%%%%%%%%%%%%%%%%%%%%%%%%%%%%%%%
\subsection{Large-Scale Environments}

\textbf{Policy Imitation.}
The imitation performances in the large-scale environments are depicted in Figure~\ref{fig:largescale} (column 1-3).
In the large-scale environments, we observe that sample efficiencies of the methods with decentralized discriminators are extremely higher than those with centralized discriminators. This is in contrast with the results in the small-scale environments in the sense that both efficiencies are comparable in those environments.
This comes from the higher variance of training centralized discriminators in the large-scale environments compared to the small-scale counterparts.
Among the methods with decentralized discriminators, we find out MA-DAAC learns much faster than the baselines.
Especially, MA-DAAC robustly achieves the best $\mathrm{NSS}$ irrespective of the number of agents, whereas the convergence of the baselines becomes slower for the increasing number of agents. 
This comes from the fact that MA-DAAC uses a shared attention-based critic as well as the off-policy samples via replay.

%%%%%%%%%%%%%%%%%%%%%%%%%
\textbf{Reward Learning.}
For the large-scale environments, we train policies and rewards over 100,000 episodes and retrain policies with MAAC and learned reward over 100,000 episodes. 
Also, the same reward models as those in the small-scale environments are used. 
The reward learning results are described in \emph{Figure~\ref{fig:smallscale}} (column 4).
Among all methods, learned rewards from MA-DAAC with decentralized discriminators leads to the best $\mathrm{NSS}$.
The performance of retrained policies decreases as the number of agents increases.

%%%%%%%%%%%%%%%%%%%%%%%%%%%%%%%%%%%%%%%%
\textbf{Number of Learnable Parameters.}
MA-DAAC is more efficient than baselines in terms of the number of parameters. As depicted in \emph{Figure~\ref{fig:largescale2}}, the number of MA-DAAC's parameters linearly increases, while the number of the baselines's parameters exponentially increase. Such an exponential increase comes from the fact that MACK doesn't share critic networks among the agents.
Note that the number of discriminator and policy parameters linearly increases for all cases. Additional results in small-scale environments are given in Appendix.

\begin{table}[t]
\caption{Imitation learning performance relative to that of MA-GAIL in \emph{Rover Tower} tasks. Note that the gain becomes much larger for the decreasing amount of available experts' demonstration. The evaluation score after training 100,000 episodes were used.}
\label{table:largescale1}
\begin{center}
\begin{tabular}{ccccc}
\toprule
\multirow{2}{*}{\# Agents}      & \multirow{2}{*}{Algorithm}    &\multicolumn{3}{c}{\# Expert Traj.}\\
\cline{3-5}
                                &                               &                          50           &                         100       &                       200 \\
\midrule
\multirow{2}{*}{8} & MA-AIRL &  $\mathrm{ 21.459 }$ &   $\mathrm{ 8.279 }$ &  $\mathbf{ 1.590 }$ \\
   & MA-DAAC &  $\mathbf{ 46.684 }$ &  $\mathbf{ 12.716 }$ &  $\mathrm{ 0.042 }$ \\
\cline{1-5}
\multirow{2}{*}{12} & MA-AIRL &   $\mathrm{ 9.285 }$ &   $\mathrm{ 8.910 }$ &  $\mathbf{ 2.400 }$ \\
   & MA-DAAC &  $\mathbf{ 44.574 }$ &  $\mathbf{ 20.154 }$ &  $\mathrm{ 1.406 }$ \\
\cline{1-5}
\multirow{2}{*}{16} & MA-AIRL &   $\mathrm{ 6.472 }$ &   $\mathrm{ 7.918 }$ &  $\mathrm{ 3.538 }$ \\
   & MA-DAAC &  $\mathbf{ 41.167 }$ &  $\mathbf{ 24.982 }$ &  $\mathbf{ 8.145 }$ \\
\bottomrule
\end{tabular}

\end{center}
\end{table}

\begin{table}[t]
\caption{Reward learning performance relative to that of MA-GAIL in \emph{Rover Tower} tasks. MA-DAAC consistently performs better than the baselines. The evaluation scores after training 100,000 episodes were used.}
\label{table:largescale2}
\begin{center}
\begin{tabular}{ccccc}
\toprule
\multirow{2}{*}{\# Agents}      & \multirow{2}{*}{Algorithm}    &\multicolumn{3}{c}{\# Expert Traj.}\\
\cline{3-5}
                                &                               &                          50           &                         100       &                       200 \\
\midrule
\multirow{2}{*}{8} & MA-AIRL &  $\mathrm{ 25.700 }$ &  $\mathrm{ 24.963 }$ &  $\mathrm{ 47.944 }$ \\
   & MA-DAAC &  $\mathbf{ 54.640 }$ &  $\mathbf{ 50.468 }$ &  $\mathbf{ 64.144 }$ \\
\cline{1-5}
\multirow{2}{*}{12} & MA-AIRL &  $\mathrm{ 21.489 }$ &  $\mathrm{ 13.177 }$ &  $\mathrm{ 25.360 }$ \\
   & MA-DAAC &  $\mathbf{ 40.887 }$ &  $\mathbf{ 36.511 }$ &  $\mathbf{ 44.920 }$ \\
\cline{1-5}
\multirow{2}{*}{16} & MA-AIRL &  $\mathrm{ 32.875 }$ &  $\mathrm{ 16.451 }$ &  $\mathrm{ 26.510 }$ \\
   & MA-DAAC &  $\mathbf{ 56.326 }$ &  $\mathbf{ 43.364 }$ &  $\mathbf{ 48.945 }$ \\
\bottomrule
\end{tabular}

\end{center}
\end{table}

%%%%%%%%%%%%%%%%%%%%%%%%%%%%%%%%%%%%%%%%%%%%%%%%%%%%%%%
\subsection{Effect of Number of Experts' Demonstration} 

For both small-scale and large-scale environments, we vary the number of available experts' demonstration among 50, 100, 200 and check its effect on policy imitation and reward learning performances (See \emph{Figure~\ref{fig:smallscale},~\ref{fig:largescale}}).
In the large-scale environments, the performance is highly affected by the number of experts' demonstration, whereas there's a negligible effect in the small-scale environments.
We believe this comes from the different size of joint observation-action spaces between small-scale and large-scale environments.
Specifically for the fixed amount of experts' demonstration, the \emph{effective amount} of training data -- the number of experts' demonstration relative to input dimensions -- decreases and causes discriminators to be more biased as the number of agents increases.
In the end, this leads to learning in the large-scale environments more difficult than learning in the small-scale environments.

Especially for learning with decentralized discriminators in the large-scale environments, we demonstrate that MA-DAAC performs better than the baselines with small amount of experts' demonstration (\emph{Table~\ref{table:largescale1},~\ref{table:largescale2}}).
For policy imitation (\emph{Table~\ref{table:largescale1}}), we observe the relative score of MA-DAAC becomes larger as the number of available demonstration decreases irrespective of the number of agents, whereas the relative score of MA-AIRL is much smaller than that of MA-DAAC.
This result supports that our method is much more robust than the baseline with respect to the number of experts' demonstration.
For reward learning (\emph{Table~\ref{table:largescale2}})), we again observe that the relative socre of MA-DAAC is consistently higher than that of MA-AIRL. 

%%%%%%%%%%%%%%%%%%%%
\section{Discussion}\label{sec:discussion}

\emph{Why do decentralized discriminators work well?}
Let's think about the sources of coordination that can lead to successful learning with decentralized discriminators.
The first one is centralized critics, which take joint observations and actions as inputs as used in many MARL algorithms. 
The second one is achieved by sampling experts' joint observations and actions that happened in the same time step.
Since experts' joint trajectories include information about how to coordinate at the specific time step, decentralized discriminators can be sufficiently trained toward the coordination of experts.
These two mechanisms allow decentralized discriminators to focus on local experiences, which highly reduces the input spaces and leads to better scalability.

\emph{Why don't we use observation-only decentralized discriminators?}
In single-agent AIRL~\cite{fu2018learning}, it was shown that a discriminator model that ignores action inputs can recover a reward function that is robust to changes of environment.
In the multi-agent problems, however, we find that observation-only discriminators may fail to imitate well, depending on the task.
In \emph{Cooperative Communication}, for example, the speaker's observation $o_s$ is fixed as the color of a goal landmark in an episode, i.e., $o_{s,0}=\cdots=o_{s,T-1}$ for the length $T$ of the episode, and the speaker's action (message) $a_{s,t}$ at $t$ becomes the part of the listener's successor observation $o_{l,t+1}$ at $t+1$. In such setting, if observation-only decentralized discriminators $D_s(o_s, o_s')$ and $D_l(o_l, o_l')$ are used, the speaker cannot learn how to send a correct message since $D_s$ does not include the message information $a_s$. Due to the incorrect message from the speaker, the observation of the listener becomes noisy, which results in poor performance of listener as well. On the other hand, an observation-only centralized discriminator $\bar{D}(o_s, o_l, o_s', o_l')$ does not suffer from the above issue since the centralized discriminators of both speaker and listener can exploit the full observation transition $(o_s, o_l)\rightarrow(o_s', o_l')$ and match the transition with transitions in experts' demonstration.
Such a problem, from the partial observable nature of multi-agent problem, opens a new challenge: learning reward functions that are both scalable and robust. We leave this problem to future work.

\begin{figure}[t]
\centering
\includegraphics[width=\linewidth]{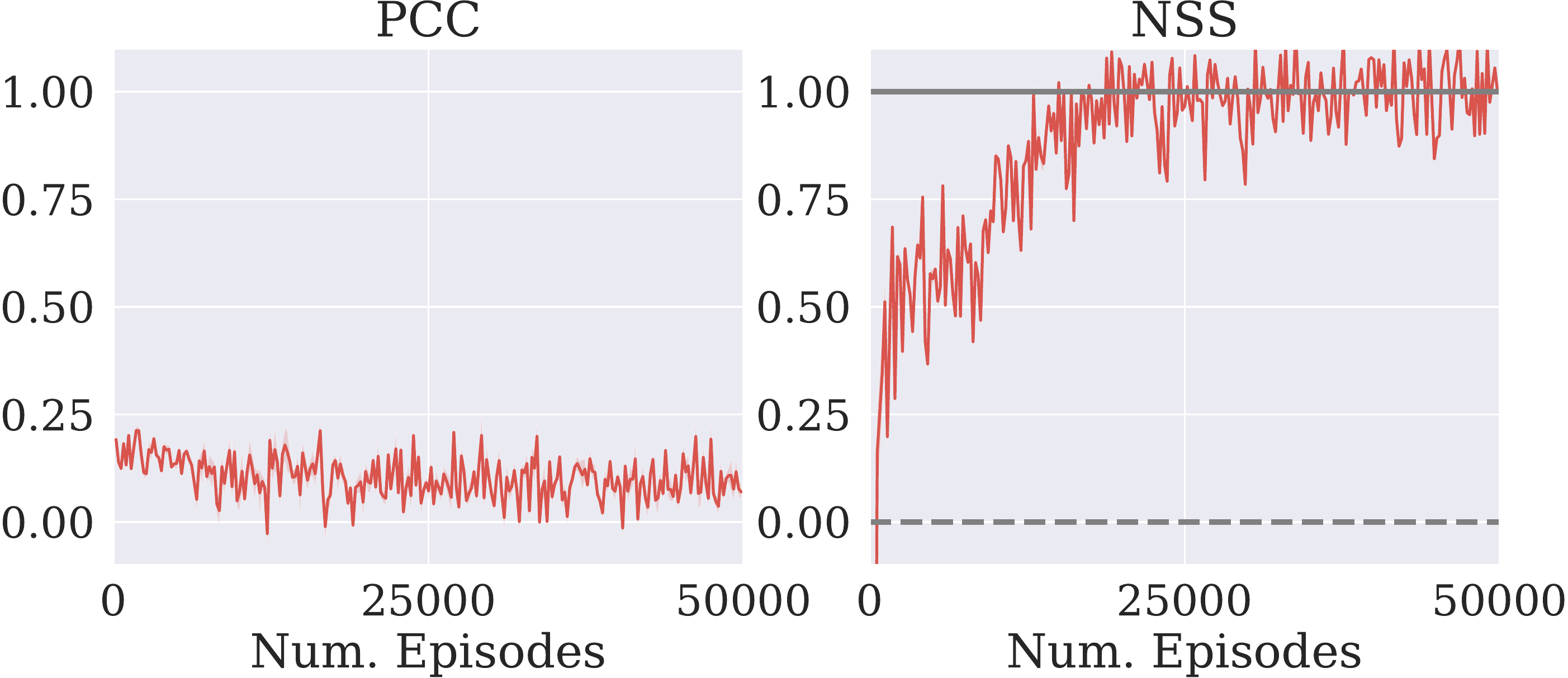}
\caption{Pearson's correlation coefficient (PCC) between ground truth rewards and learned rewards (left) and NSS (right) during retraining with learned rewards. The results imply the experts' behavior can be achieved by the learned rewards having low correlation with the ground truth.}
\label{fig:discussion}
\end{figure}

\emph{Is the correlation between learned rewards and ground truth rewards meaningful?}
In the experiments of MA-AIRL~\cite{yu2019multi}, statistical correlations between ground truth rewards -- the rewards used to train experts' policies -- and learned rewards were regarded as the performance measure of reward learning. We also check statistical correlations in our implementation, but the reward recovery performance reported in~\citet{yu2019multi} cannot be reproduced.
Discrepancies in the results may come from differences between our implementation and that of~\citet{yu2019multi}. 
However, we additionally find that high correlation between the learned rewards and the ground truth rewards is not a necessary condition of learned rewards leading to the experts' behavior, as depicted in \emph{Figure.~\ref{fig:discussion}}. 
It is well known that IRL problems are ill-defined, therefore there can exist multiple rewards that can be matched to the experts' observed trajectories.

%%%%%%%%%%%%%%%%%%%%%%%%%%%%%%%%%%%%%%%%%%
\section{Conclusion}\label{sec:conclusion}

We propose MA-DAAC, a multi-agent IRL method that is much more scalable and sample-efficient than existing works. 
We massively and rigorously analyze the performance of MA-DAAC and compare to baselines in terms of sample efficiency and the retraining score with newly defined measure ($\mathrm{NSS}$), using various types of discriminators (decentralized and centralized).
We show that MA-DAAC with decentralized discriminators outperforms other methods. 
One interesting future direction is a scalable multi-agent IRL with a centralized discriminator, so that we can efficiently interpret sequential behavior of a large number of agents by looking at the resultant centralized reward functions.

%%%%%%%%%%%%%%%%%%%%%%%%%%%%%%%%%%%%%%%%%%%%%%%%%%%%%%%%%%%%%%%%%%%%%%%%%%%%%%%%%%%%%%%%%%%%%%%%%%%%%%%%%
%% bibliography: see CFP for number of permitted pages

\bibliography{bibliography.bib}
\bibliographystyle{icml2020}

% %%%%%%%%%%%%%%%%%%%%%%%%%%%%%%%%%%%%%%%%%%%%%%%%%%%%%%%%%%%%%%%%%%%%%%%%%%%%%%%
% %%%%%%%%%%%%%%%%%%%%%%%%%%%%%%%%%%%%%%%%%%%%%%%%%%%%%%%%%%%%%%%%%%%%%%%%%%%%%%%
% % DELETE THIS PART. DO NOT PLACE CONTENT AFTER THE REFERENCES!
% %%%%%%%%%%%%%%%%%%%%%%%%%%%%%%%%%%%%%%%%%%%%%%%%%%%%%%%%%%%%%%%%%%%%%%%%%%%%%%%
% %%%%%%%%%%%%%%%%%%%%%%%%%%%%%%%%%%%%%%%%%%%%%%%%%%%%%%%%%%%%%%%%%%%%%%%%%%%%%%%
% \appendix
% \section{Do \emph{not} have an appendix here}

% \textbf{\emph{Do not put content after the references.}}
% %
% Put anything that you might normally include after the references in a separate
% supplementary file.

% We recommend that you build supplementary material in a separate document.
% If you must create one PDF and cut it up, please be careful to use a tool that
% doesn't alter the margins, and that doesn't aggressively rewrite the PDF file.
% pdftk usually works fine. 

% \textbf{Please do not use Apple's preview to cut off supplementary material.} In
% previous years it has altered margins, and created headaches at the camera-ready
% stage. 
% %%%%%%%%%%%%%%%%%%%%%%%%%%%%%%%%%%%%%%%%%%%%%%%%%%%%%%%%%%%%%%%%%%%%%%%%%%%%%%%
% %%%%%%%%%%%%%%%%%%%%%%%%%%%%%%%%%%%%%%%%%%%%%%%%%%%%%%%%%%%%%%%%%%%%%%%%%%%%%%%

\onecolumn
\icmltitle{
Scalable Multi-Agent Inverse Reinforcement Learning via Actor-Attention-Critic
\\
(Supplementary Material)
}

\appendix

%%%%%%%%%%%%%%%
\section{Details of Experiments}

\textbf{Tasks.} In our experiments, we consider four tasks built on OpenAI's Multi-Agent Particle Environments~\cite{mordatch2017emergence,lowe2017multi,iqbal2019actor}; \emph{Keep Away}, \emph{Cooperative Communication}, \emph{Cooperative Navigation}, \emph{Rover Tower}. For all tasks, the length of each episode is set to be 25.
For \emph{Rover Tower}, the number of agents is chosen among 8, 12, 16.

\textbf{MARL implementations.} We implement both MACK~\cite{song2018multi} and MAAC~\cite{iqbal2019actor} in PyTorch~\cite{PyTorch} by refactoring the codes released by the respective authors\footnote{
\url{https://github.com/ermongroup/multiagent-gail}} \footnote{
\url{https://github.com/shariqiqbal2810/MAAC}
}.
Especially to implement MACK in PyTorch, we use generalized advantage estimation (GAE)~\cite{schulman2015high} and KFAC optimizer from ACKTR implementation in PyTorch\footnote{
\url{https://github.com/ikostrikov/pytorch-a2c-ppo-acktr-gail}
}. Additionally, we apply advantage normalization technique used in OpenAI baselines\footnote{
\url{https://github.com/openai/baselines}
} which has not been implemented in original MACK implementation but highly stabilizes and improves the performance of MACK. For all policies and the value baselines in MACK, we use two-layer neural networks with 128 hidden units and LeakyReLU activations. 
For attention critic in MAAC, we use the same network architecture used in the released code. For all methods, we divide the rewards with the length of episodes.

\textbf{Experts.} We use our MAAC implementation to train experts' policies for 50,000 episodes. We use the normalization of inputs and rewards for each agent based on the methods of calculating running mean and standard deviation\footnote{
\url{https://github.com/openai/baselines/blob/master/baselines/common/running_mean_std.py}
} in OpenAI baselines. Other hyperparamters are summarized in \emph{Table~\ref{table:appendix:expert}}.
After training experts, we sample 500 episodes by using learned experts' policies and use the average score of each agent over 500 episodes to define $\mathrm{score}_{E_i}$ in $\mathrm{NSS}$.

\begin{table}[h]
\caption{Hyperparameters for training experts with MAAC}
\label{table:appendix:expert}
\begin{center}
\begin{tabular}{c|c}
\toprule
hyperparameters                             &   value \\
\midrule
discount factor                             &   0.995 \\
buffer size                                 &   50,000 \\
policy learning rate                        &   0.001 \\
target policy update rate                   &   0.01 \\
policy entropy regularization coefficient   &   0.01 \\
critic learning rate                        &   0.001 \\
target critic update rate                   &   0.01 \\
critic gradient norm clipping               &   1.0 \\
critic loss function                        &   Huber loss \\
batch size                                  &   1,000 \\
update period                               &   100 \\
\bottomrule
\end{tabular}
\end{center}
\end{table}

\textbf{Random Agents.} We sample 500 episodes by uniformly sample actions and use the average score of each agent over 500 episodes to define $\mathrm{score}_{R_i}$ in $\mathrm{NSS}$.

\textbf{Multi-agent inverse RL.} For inverse RL, we don't use any kind of normalization as opposed to learning experts' policies. While \citet{song2018multi} and \citet{yu2019multi} have considered the policy initialization with behavioral cloning, we use randomly initialized policies as done in existing works on single-agent adversarial imitation learning~\cite{ho2016generative,kostrikov2018discriminator,sasaki2018sample}.
Although we haven't demonstrated the effect of behavioral-cloning initialization on scalablity, we guess such initialization does not give significant gain to either MA-GAIL or MA-AIRL because of both using MACK as MARL algorithm.
For the discriminators of MA-GAIL, we use two-layer neural networks with 128 hidden units and LeakyReLU activation to model $D_i$. For the discriminators of MA-AIRL and MA-DAAC, we use two-layer neural networks with 128 hidden units and LeakyReLU activation for both reward estimation $g_i$ and potential shaping function $h_i$.
For the training of discriminators, we use entropy regularization of discriminators used in~\citet{ho2016generative}\footnote{
\url{https://github.com/openai/imitation}
} instead of using L2 regularization of discriminators in the released codes\footnote{
\url{https://github.com/ermongroup/MA-AIRL}
}.
This empirically leads to much better imitation and reward learning performances than using L2 regularization.
Other hyperparameters for MA-GAIL and MA-AIRL are in \emph{Table~\ref{table:appendix:baselines}}, and those for MA-DAAC are in \emph{Table~\ref{table:appendix:ours}}.

\begin{table}[h]
\caption{Hyperparameters for both MA-GAIL and MA-AIRL}
\label{table:appendix:baselines}
\begin{center}
\begin{tabular}{c|c}
\toprule
hyperparameters                             &   value \\
\midrule
$\lambda$ for GAE                           &   0.95 \\
discount factor                             &   0.995 \\
policy learning rate                        &   0.001 \\
policy target update rate                   &   0.0005 \\
policy entropy regularization coefficient   &   0.01 \\
critic learning rate                        &   0.001 \\
critic target update rate                   &   0.001 \\
critic gradient norm clipping               &   10 \\
critic loss function                        &   Huber loss \\
discriminator learning rate                 &   0.01 \\
discriminator entropy regularization coefficient & 0.01 \\
discriminator gradient norm clipping        &   10 \\
batch size                                  &   1,000 \\
\bottomrule
\end{tabular}
\end{center}
\end{table}

\begin{table}[h]
\caption{Hyperparameters for MA-DAAC}
\label{table:appendix:ours}
\begin{center}
\begin{tabular}{c|c}
\toprule
hyperparameters                             &   value \\
\midrule
discount factor                             &   0.995 \\
buffer size                                 &   1,250,000 \\
policy learning rate                        &   0.001 \\
target policy update rate                   &   0.0005 \\
policy entropy regularization coefficient   &   0.01 \\
critic learning rate                        &   0.001 \\
target critic update rate                   &   0.0005 \\
critic gradient norm clipping               &   1.0 \\
critic loss function                        &   Huber loss \\
discriminator learning rate                 &   0.0005 \\
discriminator entropy regularization coefficient & 0.01 \\
discriminator gradient norm clipping        &   10 \\
batch size                                  &   1,000 \\
update period                               &   100 \\
\bottomrule
\end{tabular}
\end{center}
\end{table}

\newpage
%%%%%%%%%%%%%%%%%%%%%%%%%%%%%%%%%%%%%%%%%%%%%
\section{Results in Small-Scale Environments}

The scores of learned policies are given in \emph{Table~\ref{table:appendix:small_scale:imitation}}, and the scores of policies retrained with learned rewards are given in   \emph{Table~\ref{table:appendix:small_scale:reward}}. For both cases, we train policies for 50,000 episodes.

\begin{table}[h]
\scriptsize
\caption{95\% confidence interval of scores after either imitation learning (MA-GAIL) or inverse reinforcement learning (MA-AIRL, MA-DAAC) in small-scale environments. Mean and confidence interval are calculated for 10 runs.}
\label{table:appendix:small_scale:imitation}
\begin{center}
\begin{tabular}{cccrrrr}
\toprule
    \multirow{2}{*}{\shortstack{Number of Experts' \\ Demonstration}}     
&   \multirow{2}{*}{Algorithm}   
&   \multirow{2}{*}{\shortstack{Discriminator \\ Type}}     
&   \multicolumn{2}{c}{Keep Away}                           
&   \multirow{2}{*}{\shortstack{Cooperative\\Communication}}
&   \multirow{2}{*}{\shortstack{Cooperative\\~~~~~~Navigation~~~~~~}} \\
    & & & \multicolumn{1}{c}{Pusher} & \multicolumn{1}{c}{Reacher} & & \\
\midrule
&Random&
&$\mathrm{114.401}\pm\mathrm{5.087}$
&$\mathrm{-25.910}\pm\mathrm{1.034}$
&$\mathrm{-57.196}\pm\mathrm{2.404}$
&$\mathrm{-178.575}\pm\mathrm{4.108}$\\
&Expert&
&$\mathrm{44.449 }\pm\mathrm{1.673}$
&$\mathrm{-9.688}\pm\mathrm{0.331}$
&$\mathrm{-15.290}\pm\mathrm{0.811}$
&$\mathrm{-77.129}\pm\mathrm{1.766}$\\
\midrule
\multirow{6}{*}{50} & \multirow{2}{*}{MA-GAIL} & Decentralized &  $\mathrm{ 44.361 }\pm\mathrm{ 3.900 }$ &  $\mathrm{ -9.580 }\pm\mathrm{ 0.784 }$ &  $\mathrm{ -17.669 }\pm\mathrm{ 1.714 }$ &  $\mathrm{ -81.393 }\pm\mathrm{ 4.447 }$ \\
    &         & Centralized &  $\mathrm{ 45.394 }\pm\mathrm{ 2.401 }$ &  $\mathrm{ -9.785 }\pm\mathrm{ 0.499 }$ &  $\mathrm{ -17.707 }\pm\mathrm{ 1.337 }$ &  $\mathrm{ -86.585 }\pm\mathrm{ 4.266 }$ \\
\cline{2-7}
    & \multirow{2}{*}{MA-AIRL} & Decentralized &  $\mathrm{ 44.515 }\pm\mathrm{ 2.736 }$ &  $\mathrm{ -9.625 }\pm\mathrm{ 0.557 }$ &  $\mathrm{ -17.305 }\pm\mathrm{ 1.354 }$ &  $\mathrm{ -79.659 }\pm\mathrm{ 4.415 }$ \\
    &         & Centralized &  $\mathrm{ 43.374 }\pm\mathrm{ 2.898 }$ &  $\mathrm{ -9.390 }\pm\mathrm{ 0.597 }$ &  $\mathrm{ -16.499 }\pm\mathrm{ 1.594 }$ &  $\mathrm{ -79.929 }\pm\mathrm{ 5.255 }$ \\
\cline{2-7}
    & \multirow{2}{*}{MA-DAAC} & Decentralized &  $\mathrm{ 44.329 }\pm\mathrm{ 3.457 }$ &  $\mathrm{ -9.595 }\pm\mathrm{ 0.708 }$ &  $\mathrm{ -20.824 }\pm\mathrm{ 1.290 }$ &  $\mathrm{ -80.248 }\pm\mathrm{ 4.182 }$ \\
    &         & Centralized &  $\mathrm{ 44.768 }\pm\mathrm{ 3.278 }$ &  $\mathrm{ -9.675 }\pm\mathrm{ 0.679 }$ &  $\mathrm{ -20.702 }\pm\mathrm{ 1.734 }$ &  $\mathrm{ -80.826 }\pm\mathrm{ 4.077 }$ \\
\cline{1-7}
\cline{2-7}
\multirow{6}{*}{100} & \multirow{2}{*}{MA-GAIL} & Decentralized &  $\mathrm{ 43.506 }\pm\mathrm{ 3.550 }$ &  $\mathrm{ -9.450 }\pm\mathrm{ 0.728 }$ &  $\mathrm{ -14.962 }\pm\mathrm{ 1.346 }$ &  $\mathrm{ -78.626 }\pm\mathrm{ 4.375 }$ \\
    &         & Centralized &  $\mathrm{ 42.891 }\pm\mathrm{ 3.278 }$ &  $\mathrm{ -9.311 }\pm\mathrm{ 0.674 }$ &  $\mathrm{ -14.884 }\pm\mathrm{ 1.343 }$ &  $\mathrm{ -81.188 }\pm\mathrm{ 4.346 }$ \\
\cline{2-7}
    & \multirow{2}{*}{MA-AIRL} & Decentralized &  $\mathrm{ 43.594 }\pm\mathrm{ 2.594 }$ &  $\mathrm{ -9.468 }\pm\mathrm{ 0.532 }$ &  $\mathrm{ -14.790 }\pm\mathrm{ 1.458 }$ &  $\mathrm{ -78.368 }\pm\mathrm{ 4.308 }$ \\
    &         & Centralized &  $\mathrm{ 43.351 }\pm\mathrm{ 2.789 }$ &  $\mathrm{ -9.404 }\pm\mathrm{ 0.575 }$ &  $\mathrm{ -15.085 }\pm\mathrm{ 1.502 }$ &  $\mathrm{ -77.788 }\pm\mathrm{ 4.194 }$ \\
\cline{2-7}
    & \multirow{2}{*}{MA-DAAC} & Decentralized &  $\mathrm{ 43.267 }\pm\mathrm{ 3.137 }$ &  $\mathrm{ -9.381 }\pm\mathrm{ 0.648 }$ &  $\mathrm{ -18.184 }\pm\mathrm{ 1.419 }$ &  $\mathrm{ -78.002 }\pm\mathrm{ 4.123 }$ \\
    &         & Centralized &  $\mathrm{ 42.303 }\pm\mathrm{ 2.838 }$ &  $\mathrm{ -9.204 }\pm\mathrm{ 0.586 }$ &  $\mathrm{ -17.542 }\pm\mathrm{ 1.504 }$ &  $\mathrm{ -79.334 }\pm\mathrm{ 4.360 }$ \\
\cline{1-7}
\cline{2-7}
\multirow{6}{*}{200} & \multirow{2}{*}{MA-GAIL} & Decentralized &  $\mathrm{ 44.626 }\pm\mathrm{ 3.028 }$ &  $\mathrm{ -9.676 }\pm\mathrm{ 0.625 }$ &  $\mathrm{ -14.508 }\pm\mathrm{ 1.520 }$ &  $\mathrm{ -78.503 }\pm\mathrm{ 4.431 }$ \\
    &         & Centralized &  $\mathrm{ 42.246 }\pm\mathrm{ 3.683 }$ &  $\mathrm{ -9.192 }\pm\mathrm{ 0.760 }$ &  $\mathrm{ -14.526 }\pm\mathrm{ 1.562 }$ &  $\mathrm{ -80.003 }\pm\mathrm{ 4.850 }$ \\
\cline{2-7}
    & \multirow{2}{*}{MA-AIRL} & Decentralized &  $\mathrm{ 42.611 }\pm\mathrm{ 4.360 }$ &  $\mathrm{ -9.268 }\pm\mathrm{ 0.883 }$ &  $\mathrm{ -14.757 }\pm\mathrm{ 1.405 }$ &  $\mathrm{ -78.825 }\pm\mathrm{ 4.465 }$ \\
    &         & Centralized &  $\mathrm{ 43.150 }\pm\mathrm{ 3.105 }$ &  $\mathrm{ -9.364 }\pm\mathrm{ 0.636 }$ &  $\mathrm{ -14.713 }\pm\mathrm{ 1.587 }$ &  $\mathrm{ -78.403 }\pm\mathrm{ 4.805 }$ \\
\cline{2-7}
    & \multirow{2}{*}{MA-DAAC} & Decentralized &  $\mathrm{ 44.041 }\pm\mathrm{ 4.201 }$ &  $\mathrm{ -9.558 }\pm\mathrm{ 0.850 }$ &  $\mathrm{ -15.820 }\pm\mathrm{ 1.292 }$ &  $\mathrm{ -77.474 }\pm\mathrm{ 4.820 }$ \\
    &         & Centralized &  $\mathrm{ 41.492 }\pm\mathrm{ 2.818 }$ &  $\mathrm{ -9.045 }\pm\mathrm{ 0.570 }$ &  $\mathrm{ -15.885 }\pm\mathrm{ 1.270 }$ &  $\mathrm{ -79.252 }\pm\mathrm{ 4.696 }$ \\
\bottomrule
\end{tabular}
\end{center}
\end{table}

\begin{table}[h]
\scriptsize
\caption{95\% confidence interval of scores after retraining with either discriminator (MA-GAIL) or learned reward functions (MA-AIRL, MA-DAAC) in small-scale environments. Mean and confidence interval are calculated for 10 runs.}
\label{table:appendix:small_scale:reward}
\begin{center}
\begin{tabular}{cccrrrr}
\toprule
    \multirow{2}{*}{\shortstack{Number of Experts' \\ Demonstration}}     
&   \multirow{2}{*}{Algorithm}   
&   \multirow{2}{*}{\shortstack{Discriminator \\ Type}}      
&   \multicolumn{2}{c}{Keep Away}                           
&   \multirow{2}{*}{\shortstack{Cooperative\\Communication}}
&   \multirow{2}{*}{\shortstack{Cooperative\\~~~~~~Navigation~~~~~~}} \\
    & & & \multicolumn{1}{c}{Pusher} & \multicolumn{1}{c}{Reacher} & & \\
\midrule
&Random&
&$\mathrm{114.401}\pm\mathrm{5.087}$
&$\mathrm{-25.910}\pm\mathrm{1.034}$
&$\mathrm{-57.196}\pm\mathrm{2.404}$
&$\mathrm{-178.575}\pm\mathrm{4.108}$\\
&Expert&
&$\mathrm{44.449 }\pm\mathrm{1.673}$
&$\mathrm{-9.688}\pm\mathrm{0.331}$
&$\mathrm{-15.290}\pm\mathrm{0.811}$
&$\mathrm{-77.129}\pm\mathrm{1.766}$\\
\midrule
\multirow{6}{*}{50} & \multirow{2}{*}{MA-GAIL} & Decentralized &    $\mathrm{ 50.302 }\pm\mathrm{ 7.254 }$ &  $\mathrm{ -10.751 }\pm\mathrm{ 1.514 }$ &  $\mathrm{ -35.746 }\pm\mathrm{ 6.407 }$ &   $\mathrm{ -88.737 }\pm\mathrm{ 10.900 }$ \\
    &         & Centralized &  $\mathrm{ 104.775 }\pm\mathrm{ 34.312 }$ &  $\mathrm{ -22.669 }\pm\mathrm{ 7.084 }$ &  $\mathrm{ -39.449 }\pm\mathrm{ 3.679 }$ &  $\mathrm{ -134.150 }\pm\mathrm{ 12.031 }$ \\
\cline{2-7}
    & \multirow{2}{*}{MA-AIRL} & Decentralized &    $\mathrm{ 42.349 }\pm\mathrm{ 1.817 }$ &   $\mathrm{ -9.112 }\pm\mathrm{ 0.366 }$ &  $\mathrm{ -26.585 }\pm\mathrm{ 3.509 }$ &    $\mathrm{ -80.355 }\pm\mathrm{ 2.471 }$ \\
    &         & Centralized &    $\mathrm{ 51.765 }\pm\mathrm{ 9.712 }$ &  $\mathrm{ -11.068 }\pm\mathrm{ 1.950 }$ &  $\mathrm{ -40.513 }\pm\mathrm{ 7.691 }$ &  $\mathrm{ -148.842 }\pm\mathrm{ 14.124 }$ \\
\cline{2-7}
    & \multirow{2}{*}{MA-DAAC} & Decentralized &    $\mathrm{ 41.520 }\pm\mathrm{ 1.304 }$ &   $\mathrm{ -8.933 }\pm\mathrm{ 0.267 }$ &  $\mathrm{ -22.827 }\pm\mathrm{ 1.158 }$ &    $\mathrm{ -72.671 }\pm\mathrm{ 0.393 }$ \\
    &         & Centralized &    $\mathrm{ 43.004 }\pm\mathrm{ 2.197 }$ &   $\mathrm{ -9.245 }\pm\mathrm{ 0.439 }$ &  $\mathrm{ -26.399 }\pm\mathrm{ 1.282 }$ &   $\mathrm{ -129.666 }\pm\mathrm{ 5.945 }$ \\
\cline{1-7}
\cline{2-7}
\multirow{6}{*}{100} & \multirow{2}{*}{MA-GAIL} & Decentralized &   $\mathrm{ 56.646 }\pm\mathrm{ 21.924 }$ &  $\mathrm{ -12.175 }\pm\mathrm{ 4.573 }$ &  $\mathrm{ -26.587 }\pm\mathrm{ 4.296 }$ &  $\mathrm{ -108.806 }\pm\mathrm{ 18.459 }$ \\
    &         & Centralized &   $\mathrm{ 94.367 }\pm\mathrm{ 24.518 }$ &  $\mathrm{ -20.866 }\pm\mathrm{ 4.722 }$ &  $\mathrm{ -40.004 }\pm\mathrm{ 4.350 }$ &  $\mathrm{ -155.032 }\pm\mathrm{ 13.791 }$ \\
\cline{2-7}
    & \multirow{2}{*}{MA-AIRL} & Decentralized &    $\mathrm{ 43.087 }\pm\mathrm{ 1.302 }$ &   $\mathrm{ -9.275 }\pm\mathrm{ 0.265 }$ &  $\mathrm{ -17.615 }\pm\mathrm{ 0.683 }$ &    $\mathrm{ -71.304 }\pm\mathrm{ 0.646 }$ \\
    &         & Centralized &    $\mathrm{ 50.066 }\pm\mathrm{ 4.977 }$ &  $\mathrm{ -10.845 }\pm\mathrm{ 1.028 }$ &  $\mathrm{ -31.832 }\pm\mathrm{ 7.729 }$ &  $\mathrm{ -179.813 }\pm\mathrm{ 23.884 }$ \\
\cline{2-7}
    & \multirow{2}{*}{MA-DAAC} & Decentralized &    $\mathrm{ 40.895 }\pm\mathrm{ 1.199 }$ &   $\mathrm{ -8.824 }\pm\mathrm{ 0.240 }$ &  $\mathrm{ -16.470 }\pm\mathrm{ 0.304 }$ &    $\mathrm{ -70.165 }\pm\mathrm{ 0.951 }$ \\
    &         & Centralized &    $\mathrm{ 39.158 }\pm\mathrm{ 1.051 }$ &   $\mathrm{ -8.478 }\pm\mathrm{ 0.210 }$ &  $\mathrm{ -17.185 }\pm\mathrm{ 0.339 }$ &   $\mathrm{ -133.241 }\pm\mathrm{ 8.657 }$ \\
\cline{1-7}
\cline{2-7}
\multirow{6}{*}{200} & \multirow{2}{*}{MA-GAIL} & Decentralized &   $\mathrm{ 71.033 }\pm\mathrm{ 29.519 }$ &  $\mathrm{ -15.388 }\pm\mathrm{ 5.903 }$ &  $\mathrm{ -20.436 }\pm\mathrm{ 3.922 }$ &    $\mathrm{ -89.098 }\pm\mathrm{ 8.290 }$ \\
    &         & Centralized &   $\mathrm{ 65.213 }\pm\mathrm{ 16.812 }$ &  $\mathrm{ -15.052 }\pm\mathrm{ 3.115 }$ &  $\mathrm{ -36.375 }\pm\mathrm{ 3.005 }$ &  $\mathrm{ -186.348 }\pm\mathrm{ 18.565 }$ \\
\cline{2-7}
    & \multirow{2}{*}{MA-AIRL} & Decentralized &    $\mathrm{ 40.246 }\pm\mathrm{ 1.109 }$ &   $\mathrm{ -8.701 }\pm\mathrm{ 0.219 }$ &  $\mathrm{ -16.079 }\pm\mathrm{ 0.173 }$ &    $\mathrm{ -70.626 }\pm\mathrm{ 0.574 }$ \\
    &         & Centralized &    $\mathrm{ 38.603 }\pm\mathrm{ 3.140 }$ &   $\mathrm{ -8.753 }\pm\mathrm{ 0.638 }$ &  $\mathrm{ -25.498 }\pm\mathrm{ 5.444 }$ &  $\mathrm{ -208.919 }\pm\mathrm{ 17.735 }$ \\
\cline{2-7}
    & \multirow{2}{*}{MA-DAAC} & Decentralized &    $\mathrm{ 37.615 }\pm\mathrm{ 0.741 }$ &   $\mathrm{ -8.165 }\pm\mathrm{ 0.151 }$ &  $\mathrm{ -15.153 }\pm\mathrm{ 0.139 }$ &    $\mathrm{ -69.530 }\pm\mathrm{ 0.589 }$ \\
    &         & Centralized &    $\mathrm{ 38.405 }\pm\mathrm{ 0.949 }$ &   $\mathrm{ -8.351 }\pm\mathrm{ 0.188 }$ &  $\mathrm{ -15.315 }\pm\mathrm{ 0.140 }$ &  $\mathrm{ -152.142 }\pm\mathrm{ 13.646 }$ \\
\bottomrule
\end{tabular}
\end{center}
\end{table}

\newpage
%%%%%%%%%%%%%%%%%
\section{Results in Large-Scale Environments}

The scores of learned policies are given in \emph{Table~\ref{table:appendix:large_scale:imitation}}, and the scores of policies retrained with learned rewards are given in \emph{Table~\ref{table:appendix:large_scale:reward}}. For both cases, we train policies for 100,000 episodes.

\begin{table}[h]
\scriptsize
\caption{95\% confidence interval of scores after either imitation learning (MA-GAIL) or inverse reinforcement learning (MA-AIRL, MA-DAAC) in large-scale environments. Mean and confidence interval are calculated for 10 runs.}
\label{table:appendix:large_scale:imitation}
\begin{center}
\begin{tabular}{cccrrr}
\toprule
    \multirow{2}{*}{\shortstack{Number of Experts' \\ Demonstration}}     
&   \multirow{2}{*}{Algorithm}   
&   \multirow{2}{*}{\shortstack{Discriminator \\ Type}}      
&   \multicolumn{3}{c}{Rover Tower} \\
    & & & \multicolumn{1}{c}{8 Agents} & \multicolumn{1}{c}{12 Agents} & \multicolumn{1}{c}{16 Agents} \\
\midrule

&Random&
&$\mathrm{-8.350}\pm\mathrm{7.324}$
&$\mathrm{-7.691}\pm\mathrm{7.399}$
&$\mathrm{-7.016}\pm\mathrm{7.447}$\\
&Expert&
&$\mathrm{127.506}\pm\mathrm{7.136}$
&$\mathrm{126.258}\pm\mathrm{7.149}$
&$\mathrm{123.905}\pm\mathrm{7.349}$\\
\midrule
\multirow{6}{*}{50} & \multirow{2}{*}{MA-GAIL} & Decentralized &  $\mathrm{ 76.871 }\pm\mathrm{ 12.076 }$ &   $\mathrm{ 61.583 }\pm\mathrm{ 6.299 }$ &   $\mathrm{ 41.017 }\pm\mathrm{ 6.452 }$ \\
    &         & Centralized &   $\mathrm{ 22.687 }\pm\mathrm{ 6.884 }$ &    $\mathrm{ 2.050 }\pm\mathrm{ 7.660 }$ &  $\mathrm{ -42.226 }\pm\mathrm{ 6.223 }$ \\
\cline{2-6}
    & \multirow{2}{*}{MA-AIRL} & Decentralized &   $\mathrm{ 98.330 }\pm\mathrm{ 9.890 }$ &   $\mathrm{ 70.868 }\pm\mathrm{ 7.709 }$ &   $\mathrm{ 47.488 }\pm\mathrm{ 7.508 }$ \\
    &         & Centralized &   $\mathrm{ 13.471 }\pm\mathrm{ 8.119 }$ &   $\mathrm{ 15.178 }\pm\mathrm{ 7.308 }$ &   $\mathrm{ -5.344 }\pm\mathrm{ 5.054 }$ \\
\cline{2-6}
    & \multirow{2}{*}{MA-DAAC} & Decentralized &  $\mathrm{ 123.555 }\pm\mathrm{ 9.899 }$ &  $\mathrm{ 106.157 }\pm\mathrm{ 6.865 }$ &  $\mathrm{ 82.184 }\pm\mathrm{ 10.106 }$ \\
    &         & Centralized &   $\mathrm{ 27.824 }\pm\mathrm{ 9.566 }$ &   $\mathrm{ 13.148 }\pm\mathrm{ 3.904 }$ &   $\mathrm{ -5.897 }\pm\mathrm{ 2.865 }$ \\
\cline{1-6}
\cline{2-6}
\multirow{6}{*}{100} & \multirow{2}{*}{MA-GAIL} & Decentralized &  $\mathrm{ 116.679 }\pm\mathrm{ 9.450 }$ &   $\mathrm{ 99.835 }\pm\mathrm{ 3.838 }$ &   $\mathrm{ 79.035 }\pm\mathrm{ 7.221 }$ \\
    &         & Centralized &   $\mathrm{ 76.053 }\pm\mathrm{ 7.845 }$ &   $\mathrm{ 11.679 }\pm\mathrm{ 7.368 }$ &  $\mathrm{ -18.358 }\pm\mathrm{ 7.257 }$ \\
\cline{2-6}
    & \multirow{2}{*}{MA-AIRL} & Decentralized &  $\mathrm{ 124.958 }\pm\mathrm{ 9.346 }$ &  $\mathrm{ 108.745 }\pm\mathrm{ 6.382 }$ &   $\mathrm{ 86.953 }\pm\mathrm{ 7.406 }$ \\
    &         & Centralized &   $\mathrm{ 24.219 }\pm\mathrm{ 8.562 }$ &   $\mathrm{ 15.382 }\pm\mathrm{ 5.354 }$ &    $\mathrm{ 7.457 }\pm\mathrm{ 4.118 }$ \\
\cline{2-6}
    & \multirow{2}{*}{MA-DAAC} & Decentralized &  $\mathrm{ 129.395 }\pm\mathrm{ 8.851 }$ &  $\mathrm{ 119.990 }\pm\mathrm{ 3.848 }$ &  $\mathrm{ 104.017 }\pm\mathrm{ 7.968 }$ \\
    &         & Centralized &  $\mathrm{ 69.189 }\pm\mathrm{ 10.891 }$ &   $\mathrm{ 16.021 }\pm\mathrm{ 5.337 }$ &    $\mathrm{ 4.275 }\pm\mathrm{ 5.001 }$ \\
\cline{1-6}
\cline{2-6}
\multirow{6}{*}{200} & \multirow{2}{*}{MA-GAIL} & Decentralized &  $\mathrm{ 129.867 }\pm\mathrm{ 8.288 }$ &  $\mathrm{ 122.465 }\pm\mathrm{ 3.853 }$ &  $\mathrm{ 104.923 }\pm\mathrm{ 6.920 }$ \\
    &         & Centralized &   $\mathrm{ 81.646 }\pm\mathrm{ 8.043 }$ &  $\mathrm{ 41.779 }\pm\mathrm{ 11.829 }$ &    $\mathrm{ 9.384 }\pm\mathrm{ 4.712 }$ \\
\cline{2-6}
    & \multirow{2}{*}{MA-AIRL} & Decentralized &  $\mathrm{ 131.456 }\pm\mathrm{ 8.850 }$ &  $\mathrm{ 124.865 }\pm\mathrm{ 3.356 }$ &  $\mathrm{ 108.461 }\pm\mathrm{ 7.676 }$ \\
    &         & Centralized &  $\mathrm{ 55.244 }\pm\mathrm{ 10.558 }$ &   $\mathrm{ 17.093 }\pm\mathrm{ 5.807 }$ &    $\mathrm{ 9.268 }\pm\mathrm{ 4.743 }$ \\
\cline{2-6}
    & \multirow{2}{*}{MA-DAAC} & Decentralized &  $\mathrm{ 129.909 }\pm\mathrm{ 8.638 }$ &  $\mathrm{ 123.871 }\pm\mathrm{ 3.986 }$ &  $\mathrm{ 113.068 }\pm\mathrm{ 7.059 }$ \\
    &         & Centralized &   $\mathrm{ 99.000 }\pm\mathrm{ 8.405 }$ &   $\mathrm{ 31.292 }\pm\mathrm{ 5.700 }$ &    $\mathrm{ 5.894 }\pm\mathrm{ 4.343 }$ \\
\bottomrule
\end{tabular}
\end{center}
\end{table}

\begin{table}[h]
\scriptsize
\caption{95\% confidence interval of scores after retraining with either discriminator (MA-GAIL) or learned reward functions (MA-AIRL, MA-DAAC) in large-scale environments. Mean and confidence interval are calculated for 10 runs.}
\label{table:appendix:large_scale:reward}
\begin{center}
\begin{tabular}{cccrrr}
\toprule
    \multirow{2}{*}{\shortstack{Number of Experts' \\ Demonstration}}     
&   \multirow{2}{*}{Algorithm}   
&   \multirow{2}{*}{\shortstack{Discriminator \\ Type}}    
&   \multicolumn{3}{c}{Rover Tower} \\
    & & & \multicolumn{1}{c}{8 Agents} & \multicolumn{1}{c}{12 Agents} & \multicolumn{1}{c}{16 Agents} \\
\midrule

&Random&
&$\mathrm{-8.350}\pm\mathrm{7.324}$
&$\mathrm{-7.691}\pm\mathrm{7.399}$
&$\mathrm{-7.016}\pm\mathrm{7.447}$\\
&Expert&
&$\mathrm{127.506}\pm\mathrm{7.136}$
&$\mathrm{126.258}\pm\mathrm{7.149}$
&$\mathrm{123.905}\pm\mathrm{7.349}$\\
\midrule
\multirow{6}{*}{50} & \multirow{2}{*}{MA-GAIL} & Decentralized &   $\mathrm{ 47.884 }\pm\mathrm{ 6.650 }$ &   $\mathrm{ 45.212 }\pm\mathrm{ 5.701 }$ &   $\mathrm{ 20.972 }\pm\mathrm{ 4.398 }$ \\
    &         & Centralized &  $\mathrm{ -43.134 }\pm\mathrm{ 4.511 }$ &  $\mathrm{ -40.042 }\pm\mathrm{ 2.947 }$ &  $\mathrm{ -46.644 }\pm\mathrm{ 3.469 }$ \\
\cline{2-6}
    & \multirow{2}{*}{MA-AIRL} & Decentralized &   $\mathrm{ 73.584 }\pm\mathrm{ 7.715 }$ &   $\mathrm{ 66.700 }\pm\mathrm{ 4.429 }$ &   $\mathrm{ 53.848 }\pm\mathrm{ 4.561 }$ \\
    &         & Centralized &  $\mathrm{ -24.649 }\pm\mathrm{ 8.476 }$ &  $\mathrm{ -13.008 }\pm\mathrm{ 3.343 }$ &  $\mathrm{ -21.239 }\pm\mathrm{ 4.914 }$ \\
\cline{2-6}
    & \multirow{2}{*}{MA-DAAC} & Decentralized &  $\mathrm{ 102.523 }\pm\mathrm{ 5.147 }$ &   $\mathrm{ 86.099 }\pm\mathrm{ 3.216 }$ &   $\mathrm{ 77.299 }\pm\mathrm{ 2.930 }$ \\
    &         & Centralized &   $\mathrm{ 37.058 }\pm\mathrm{ 5.255 }$ &   $\mathrm{ -3.786 }\pm\mathrm{ 3.045 }$ &   $\mathrm{ -0.402 }\pm\mathrm{ 4.640 }$ \\
\cline{1-6}
\cline{2-6}
\multirow{6}{*}{100} & \multirow{2}{*}{MA-GAIL} & Decentralized &   $\mathrm{ 56.820 }\pm\mathrm{ 4.052 }$ &   $\mathrm{ 72.801 }\pm\mathrm{ 5.130 }$ &   $\mathrm{ 55.605 }\pm\mathrm{ 4.394 }$ \\
    &         & Centralized &  $\mathrm{ -40.101 }\pm\mathrm{ 3.222 }$ &  $\mathrm{ -33.989 }\pm\mathrm{ 4.197 }$ &  $\mathrm{ -40.653 }\pm\mathrm{ 4.001 }$ \\
\cline{2-6}
    & \multirow{2}{*}{MA-AIRL} & Decentralized &   $\mathrm{ 81.783 }\pm\mathrm{ 5.797 }$ &   $\mathrm{ 85.978 }\pm\mathrm{ 3.853 }$ &   $\mathrm{ 72.056 }\pm\mathrm{ 2.421 }$ \\
    &         & Centralized &  $\mathrm{ -38.940 }\pm\mathrm{ 6.682 }$ &  $\mathrm{ -20.038 }\pm\mathrm{ 5.944 }$ &  $\mathrm{ -19.045 }\pm\mathrm{ 4.746 }$ \\
\cline{2-6}
    & \multirow{2}{*}{MA-DAAC} & Decentralized &  $\mathrm{ 107.288 }\pm\mathrm{ 3.747 }$ &  $\mathrm{ 109.312 }\pm\mathrm{ 3.325 }$ &   $\mathrm{ 98.969 }\pm\mathrm{ 2.906 }$ \\
    &         & Centralized &   $\mathrm{ 49.207 }\pm\mathrm{ 5.798 }$ &   $\mathrm{ -6.471 }\pm\mathrm{ 2.168 }$ &    $\mathrm{ 2.674 }\pm\mathrm{ 3.186 }$ \\
\cline{1-6}
\cline{2-6}
\multirow{6}{*}{200} & \multirow{2}{*}{MA-GAIL} & Decentralized &   $\mathrm{ 44.758 }\pm\mathrm{ 8.127 }$ &   $\mathrm{ 72.653 }\pm\mathrm{ 3.115 }$ &   $\mathrm{ 57.997 }\pm\mathrm{ 1.997 }$ \\
    &         & Centralized &  $\mathrm{ -24.300 }\pm\mathrm{ 3.624 }$ &  $\mathrm{ -25.582 }\pm\mathrm{ 7.643 }$ &  $\mathrm{ -38.050 }\pm\mathrm{ 3.034 }$ \\
\cline{2-6}
    & \multirow{2}{*}{MA-AIRL} & Decentralized &   $\mathrm{ 92.702 }\pm\mathrm{ 2.447 }$ &   $\mathrm{ 98.013 }\pm\mathrm{ 3.546 }$ &   $\mathrm{ 84.507 }\pm\mathrm{ 2.464 }$ \\
    &         & Centralized &  $\mathrm{ -48.468 }\pm\mathrm{ 8.504 }$ &  $\mathrm{ -29.585 }\pm\mathrm{ 6.222 }$ &  $\mathrm{ -21.437 }\pm\mathrm{ 4.652 }$ \\
\cline{2-6}
    & \multirow{2}{*}{MA-DAAC} & Decentralized &  $\mathrm{ 108.902 }\pm\mathrm{ 4.831 }$ &  $\mathrm{ 117.573 }\pm\mathrm{ 1.950 }$ &  $\mathrm{ 106.942 }\pm\mathrm{ 2.395 }$ \\
    &         & Centralized &   $\mathrm{ 75.090 }\pm\mathrm{ 3.188 }$ &    $\mathrm{ 5.999 }\pm\mathrm{ 3.978 }$ &    $\mathrm{ 2.741 }\pm\mathrm{ 3.455 }$ \\
\bottomrule
\end{tabular}
\end{center}
\end{table}

\newpage
%%%%%%%%%%%%%%%%%%%%%%%%%%%%%%%%%%%%%%%%%%%%%%%%%%%%%%%%%%%%%%%%%%%%%%%%%%%%%%%%%%%%%%%%%%%%%%%%%%%%%%%%%
%% bibliography: see CFP for number of permitted pages

% %%%%%%%%%%%%%%%%%%%%%%%%%%%%%%%%%%%%%%%%%%%%%%%%%%%%%%%%%%%%%%%%%%%%%%%%%%%%%%%
% %%%%%%%%%%%%%%%%%%%%%%%%%%%%%%%%%%%%%%%%%%%%%%%%%%%%%%%%%%%%%%%%%%%%%%%%%%%%%%%
% % DELETE THIS PART. DO NOT PLACE CONTENT AFTER THE REFERENCES!
% %%%%%%%%%%%%%%%%%%%%%%%%%%%%%%%%%%%%%%%%%%%%%%%%%%%%%%%%%%%%%%%%%%%%%%%%%%%%%%%
% %%%%%%%%%%%%%%%%%%%%%%%%%%%%%%%%%%%%%%%%%%%%%%%%%%%%%%%%%%%%%%%%%%%%%%%%%%%%%%%
% \appendix
% \section{Do \emph{not} have an appendix here}

% \textbf{\emph{Do not put content after the references.}}
% %
% Put anything that you might normally include after the references in a separate
% supplementary file.

% We recommend that you build supplementary material in a separate document.
% If you must create one PDF and cut it up, please be careful to use a tool that
% doesn't alter the margins, and that doesn't aggressively rewrite the PDF file.
% pdftk usually works fine. 

% \textbf{Please do not use Apple's preview to cut off supplementary material.} In
% previous years it has altered margins, and created headaches at the camera-ready
% stage. 
% %%%%%%%%%%%%%%%%%%%%%%%%%%%%%%%%%%%%%%%%%%%%%%%%%%%%%%%%%%%%%%%%%%%%%%%%%%%%%%%
% %%%%%%%%%%%%%%%%%%%%%%%%%%%%%%%%%%%%%%%%%%%%%%%%%%%%%%%%%%%%%%%%%%%%%%%%%%%%%%%

\end{document}